\documentstyle [12pt] {article}
\makeatletter
\ifcase \@ptsize
\or % mods for 11 pt
\or % mods for 12 pt
\fi
\advance\textheight by \topskip

\ifcase \@ptsize
\or % mods for 11 pt
\or % mods for 12 pt
\fi

\def\@citex[#1]#2{%
\if@filesw \immediate \write \@auxout {\string \citation {#2}}\fi
\@tempcntb\m@ne \let\@h@ld\relax \def\@citea{}%
\@cite{%
  \@for \@citeb:=#2\do {%
    \@ifundefined {b@\@citeb}%
      {\@h@ld\@citea\@tempcntb\m@ne{\bf ?}%
      \@warning {Citation `\@citeb ' on page \thepage \space undefined}}%
      {\@tempcnta\@tempcntb \advance\@tempcnta\@ne%
      \@tempcntb\number\csname b@\@citeb \endcsname \relax%
      \ifnum\@tempcnta=\@tempcntb %   Number follows previous--hold on to it
        \ifx\@h@ld\relax%
          \edef \@h@ld{\@citea\csname b@\@citeb\endcsname}%
        \else%
          \edef\@h@ld{\ifmmode{-}\else--\fi\csname b@\@citeb\endcsname}%
        \fi%
      \else%   %  non-successor--dump what's held and do this one
        \@h@ld\@citea\csname b@\@citeb \endcsname%
        \let\@h@ld\relax%
      \fi}%
    \def\@citea{,\penalty\@highpenalty\,}%
  }\@h@ld%
}{#1}}

\makeatother
\begin{document}
\hfuzz=100pt
\textheight 24.0cm
\topmargin -0.5in
%\baselineskip 16pt
%\parskip 18pt
%\parindent 30pt
%\def\mc{\,\raise -2.truept\hbox{\rlap{\hbox{$\sim$}}\raise5.truept
%\hbox{$<$}\ }}
%\def\Mc{\,\raise -2.truept\hbox{\rlap{\hbox{$\sim$}}\raise5.truept
%\hbox{$>$}\ }}%
%
%
%%%%%%%%%%%%%%%%%%%%%%%%%%%%%%%%%%%%%%%%%%%%%%%%%%%%%%%%%%%%%%%
%
%    List of  the     commands
%%%%%%%%%%%%%%%%%%%%%%%%%%%%%%%%%%%%%%%%%%%%%%%%%%%%%%%%%%%%%%%
%
\newcommand{\be}{\begin{equation}}
\newcommand{\ee}{\end{equation}}
\newcommand{\bea}{\begin{eqnarray}}
\newcommand{\eea}{\end{eqnarray}}
\begin{titlepage}
\makeatletter
\def \thefootnote {\fnsymbol {footnote}} \def \@makefnmark {
\hbox to 0pt{$^{\@thefnmark }$\hss }}
\makeatother
\begin{flushright}
SHEP 94/95--17\\
December, 1994\\
hep-th/9412133
\end{flushright}
\vspace{1.5cm}
\begin{center}
{ \Large \bf Fractional Supersymmetry}\\

\vspace{2cm}
%{\large\bf H. Arfaei} \footnote{e-mail: arfaei@IREARN.BITNET }\\
%\vspace{.5cm}
%Department of Physics\\
%Sharif University of Technology\\
%P. O. Box 11365-9161\\
%Tehran, Iran\\
%\vspace{.5cm}
%and\\
%\vspace{.5cm}
{\large\bf Noureddine Mohammedi}
\footnote{e-mail: nouri@hep.ph.soton.ac.uk}
%\footnote
%{Work supported by the Alexander von Humboldt-Stiftung.}
\\
\vspace{.5cm}
Department of Physics\\
University of Southampton \\
Highfield\\
Southampton SO17 1BJ \\
U.K.\\

\baselineskip 18pt
%\vspace{.2in}
\vspace{1cm}
{\large\bf Abstract}
\end{center}
A symmetry between bosonic coordinates and some Grassmannian-type
coordinates is presented. Commuting two of these Grassmannian-type
variables results in an arbitrary phase factor (not
just a minus sign). This symmetry is also realised at the level
of the field theory.\\
\setcounter {footnote}{0}
\end{titlepage}
\baselineskip 20pt
%

%\setcounter{chapter}{1}
%\setcounter{section}{1}
%\setcounter{subsection}{1}
%\section{Introduction}
%\section{}

Supersymmetry is a symmetry between bosons and fermions [1,2], where the
fermions pick up a minus sign (or a phase factor of $e^{i\pi}$) each
time two of them commute. The aim of this note is to generalise
the ideas of supersymmetry. We find a symmetry between bosons and
Grassmannian-type variables.  The Grassmannian-type variables
give rise to an arbitrary phase factor each time two of them
cross each other. We believe that this symmetry describes
particles  with fractional (or exotic) statistics and could be
of interest in solving problems in solid state physics such as
fractional Hall effect [3] and anyon superconductivity [4]. The
analyses carried out here could also be extended to the
theory of differential calculus on quantum spaces and
quantum groups developed in refs.[5--10].
\par
Let us suppose that we have some variables $\theta_i$,
$i=1,\dots,d$, which satisfy
the commutation relations [9,10,11]
\be
\theta_i\theta_j=q\theta_j\theta_i \,\,\,\,,\,\,\,\, i<j\,\,\,\,,
\ee
where $q$ is a complex number. We also define differentiation
with respect to these variables in the following way [11]
\bea
{\partial\theta_i\over\partial\theta_j}&=&\delta^j_i  \nonumber\\
{\partial\theta_i^2\over\partial\theta_i} &=&
\left(1+q \right)\theta_i\,\,\,\,.
\eea
\par
The commutation relations between derivatives and variables
are given by
\bea
{\partial\over\partial\theta_i}\theta_j&=&{1\over q}\theta_j
{\partial\over\partial\theta_i}
\,\,\,\,,\,\,\,\,i<j \nonumber\\
{\partial\over\partial\theta_i}\theta_j&=&q\theta_j
{\partial\over\partial\theta_i}
\,\,\,\,,\,\,\,\,i>j \,\,\,\,.
\eea
These commutation relations are consistent with the defining
commutation relations of the variables $\theta_i$. Furthermore,
the commutation relations among the derivatives are consistently found
to be
\be
{\partial\over\partial\theta_i}
{\partial\over\partial\theta_j}=q
{\partial\over\partial\theta_j}
{\partial\over\partial\theta_i}
\,\,\,\,,\,\,\,\,i<j \,\,\,\,\,\,.
\ee
\par
A straightforward calculation gives
\be
{\partial\theta_i^m\over\partial\theta_i}
={1-q^m\over 1-q}\theta_i^{m-1}\,\,\,\,.
\ee
Notice that if for some positive number $r$ such that $q^r=1$ then
we have ${\partial\theta_i^r\over\partial\theta_i}=0$. It would,
therefore, be plausible to postulate that [11]
\be
\theta_i^r=0 \,\,\,\,\,\, \mbox{for}\,\,\,\,q^r=1\,\,\,\,.
\ee
For $r=2$ one has the usual fermionic variables. In what
follows we always consider $r\ge 3$ and we take
\be
q=\exp\left({2\pi i\over r}\right)\,\,\,\,\,\,.
\ee
\par
In this theory we still have the usual bosonic coordinates $x_\mu$,
$\mu=1,\dots,n$, which satisfy
\be
x_\mu x_\nu=x_\nu x_\mu\,\,\,\,\,\,\,,\,\,\,\,\,\,\,\,
x_\mu\theta_i=\theta_i x_\mu\,\,\,\,\,.
\ee
Let us now explore the possible transformations on both
$\theta_i$ and $x_\mu$. The obvious transformations in the
space specified by $\theta_i$ would be translations.
These we write as
\be
\theta_i'=\theta_i + \varepsilon_i\,\,\,\,.
\ee
These transformations must respect the commutation relations
satisfied by $\theta_i$. This leads to the following
commutation relations for $\varepsilon_i$
\bea
\varepsilon_i\varepsilon_j&=&q \varepsilon_j\varepsilon_i \,\,\,\,\,
\mbox{for}\,\,\,\,\,i<j \nonumber\\
\varepsilon_i\theta_j&=&q \theta_j\varepsilon_i \,\,\,\,\,
\mbox{for}\,\,\,\,\,i<j \nonumber\\
\theta_i\varepsilon_j&=&q \varepsilon_j\theta_i \,\,\,\,\,
\mbox{for}\,\,\,\,\,i<j \,\,\,\,\,.
\eea
Furthermore we must also have $\theta_i'^r=0$, which yields
\bea
&\theta_i\varepsilon_i=q\varepsilon_i\theta_i &\nonumber\\
&\varepsilon_i^r=0\,\,\,\,.&
\eea
\par
We can construct a symmetry between $\theta_i$ and $x_\mu$ by
allocating to $x_\mu$ the transformation
\be
x_\mu'=x_\mu+\sigma_\mu^{ij}\varepsilon_i\theta_j^{r-1}\,\,\,\,.
\ee
The matrix $\sigma_\mu^{ij}$ is, for $r\ge 3$,  diagonal. The new
coordinate $x_\mu'$, however, commutes with $\theta_i$ and
$\varepsilon_i$ only when
\be
\varepsilon_i^2=0\,\,\,\,\,\,.
\ee
This last equation, obviously, leads to $\varepsilon_i^r=0$ for
$r\ge 3$.
\par
A superfield $\Phi\left(x^\mu,\theta_i\right)$ is a function
that is expandable as a Taylor series in the coordinates
$\theta_i$ [12,13]. For an infinitesimal transformation $\varepsilon_i$
we have
\be
\Phi\left(x'^\mu,\theta_i'\right)=
\Phi\left(x^\mu,\theta_i\right) + \varepsilon_j Q_j
\Phi\left(x^\mu,\theta_i\right)\,\,\,\,\,,
\ee
where the supersymmetric charge is given by
\be
Q_i={\partial\over\partial\theta_i}+\sigma_\mu^{ij}\theta_j^{r-1}
{\partial\over\partial x_\mu}\,\,\,\,.
\ee
\par
The corresponding covariant derivative is found to be given by
\be
D_i={\partial\over\partial\theta_i}
+{1\over q}\sigma_\mu^{ij}\theta_j^{r-1}
{\partial\over\partial x_\mu}\,\,\,\,.
\ee
This is a covariant derivative in the sense that
\be
{\partial\over\partial\theta_i'}
+{1\over q}\sigma_\mu^{ij}\theta_j'^{r-1}
{\partial\over\partial x'_\mu}
={\partial\over\partial\theta_i}
+{1\over q}\sigma_\mu^{ij}\theta_j^{r-1}
{\partial\over\partial x_\mu}\,\,\,\,.
\ee
The algebra satisfied by the supersymmetric charges is
\be
Q_kQ_l=qQ_lQ_k\,\,\,\,\,,\,\,\,\,\, k<l\,\,\,\,\,\,.
\ee
Similarly, the commutation relations between the covariant
derivatives $D_i$ are also given by
\be
D_kD_l=qD_lD_k\,\,\,\,\,,\,\,\,\,\, k<l\,\,\,\,\,\,.
\ee
The third commutation relations are between the covariant
derivatives $D_i$ and the supersymmetric charges $Q_i$.
These are found to be
\bea
Q_kD_l&=&qD_lQ_k\,\,\,\,\,,\,\,\,\,\, k<l\,\,\,\,\,\,
\nonumber\\
Q_kD_l&=&{1\over q}D_lQ_k\,\,\,\,\,,\,\,\,\,\, k>l\,\,\,\,\,\,.
\eea
\par
The powers of the supersymmetric charge are computed using the formula
\be
Q_i^s={\partial^s\over\partial\theta_i^s}+
\sum_{p=0}^{s-1}A_p^{(s)}\theta_i^{r-s+p}\sigma_\mu^{ii}
{\partial^p\over\partial\theta_i^p}
{\partial\over\partial x_\mu}
+\dots\,\,\,\,.
\ee
Here the dots stand for terms that vanish for
$s\le r$. The coefficients
$A_p^{(s)}$ are given by the recursion relations
\bea
A_p^{(s+1)}=\delta^s_p + S(r-s+p)A^{(s)}_p +q^{r-s+p-1}A^{(s)}_{p-1}
\nonumber\\
0\le p\le s \,\,\,,\,\,\,A_{-1}^{(s)}=0\,\,\,\,,
A_0^{(1)}=1\,\,\,\,.
\eea
The function $S$ is defined as
\be
S(x)={1-q^x\over 1-q}\,\,\,\,.
\ee
Taking $q=\exp(2\pi i/r)$, we find
\be
Q_i^r=S(r-1)S(r-2)\dots S(1)\sigma^{ii}_\mu
{\partial\over\partial x_\mu}\,\,\,\,.
\ee
\par
The general expansion of a superfield $\Phi\left(x,\theta\right)$
is given by
\be
\Phi\left(x,\theta\right)=\theta_1^{n_1}\dots\theta_d^{n_d}
B_{n_{1}\dots n_{d}}\left(x\right)\,\,\,\,\,\,,\,\,\,\,\,\,
n_i=0,\dots,r-1\,\,\,\,\,\,.
\ee
The infinitesimal transformations of the components
$B_{n_{1}\dots n_{d}}$ are given, for a fixed $i$, by
\bea
\delta_{\varepsilon_{i}}B_{n_{1}\dots,n_{i-1}, n_{i}-1,n_{i+1}
\dots n_{d}}&=&q^{(1-n_{i}+n_{i+1}+\dots+n_{d})}S(n_{i})
\varepsilon_{i}B_{n_{1}\dots n_{i}\dots n_{d}}
\,\,\,\,\,\,\,\, \mbox{for} \,\,\,\,\,\,\, n_i\ne r-1
\nonumber\\
\delta_{\varepsilon_{i}}B_{n_{1}\dots,n_{i-1}, r-1,n_{i+1}
\dots n_{d}}&=&q^{(1+n_{i+1}+\dots+n_{d})}\sigma^{ii}_\mu
\varepsilon_{i}{\partial\over\partial x_\mu}
B_{n_{1}\dots n_{i-1},0,n_{i+1}\dots n_{d}}
\,\,\,\,\,.
\eea
The construction of a supersymmetric action follows from the fact
that the transformation of the component $B_{r-1,\dots, r-1}$ is
a spacetime divergence. On the other hand this component is
given by
\bea
B_{r-1,\dots,r-1}\left(x\right)&=&aD^{r-1}_{d}\dots D^{r-1}_1
\left(\Phi(x,\theta)\right)|_{\theta=0}
\nonumber\\
a&=&\left[S(r-1)S(r-2)\dots S(1)\right]^d
\,\,\,\,.
\eea
Therefore a supersymmetric action is found by taking
the spacetime integral of the component $B_{r-1,\dots,r-1}$.
This action is written as
\bea
S&=&{1\over a}\int {\mbox d}^nxD^{r-1}_d\dots
D^{r-1}_1\left(\Phi(x,\theta)\right)
|_{\theta=0}\nonumber\\
&=&{1\over a}\int {\mbox d}^nxD^{r-1}_d\dots
D^{r-1}_1\left(\Phi(x,\theta)\right)
\eea
The second equality arises because the $\theta$-dependent terms
(after the action of the derivatives) are all surface terms and do
not contribute to the integral.
\par
The applications and implications of this symmetry deserve
a thorough exploration.

\vspace{0.5cm}
\paragraph{Acknowledgements:}
I would like to thank E. Ahmed for very useful discussions and
correspondence.

\end{document}